\documentclass[10pt,twocolumn,oneside,a4paper,final]{IEEEtran}
\usepackage{indentfirst}
\usepackage{setspace}
\usepackage[numbers,sort&compress]{natbib}
\usepackage{graphicx}
\usepackage{amstext}
\usepackage{algorithm}
\usepackage{algorithmic}
\usepackage{mathrsfs}
\usepackage{amssymb}
\usepackage{amsmath}
\usepackage{epstopdf}
\usepackage{CJK}
\usepackage{stfloats}
\usepackage{url}
\usepackage{color}
\usepackage[dvipsnames]{xcolor}
\usepackage{enumerate}
\usepackage{diagbox}
\usepackage{tabularx}
\usepackage{footnote}
\makesavenoteenv{tabular}
\allowdisplaybreaks[0]
\usepackage[normalem]{ulem} 

\begin{document}
%
\title{Frequency-Mixing Intelligent Reflecting Surfaces for Nonlinear Wireless Propagation}
%
%
%


\author{Jide Yuan,~\IEEEmembership{Member,~IEEE}, Elisabeth De Carvalho,~\IEEEmembership{Senior Member,~IEEE}, Robin Jess Williams,~\IEEEmembership{Member,~IEEE}, Emil Bj\"{o}rnson,~\IEEEmembership{Senior Member,~IEEE}
and Petar Popovski,~\IEEEmembership{Fellow,~IEEE}
}

\maketitle

\begin{abstract}


We introduce the concept of frequency-mixing intelligent reflecting surface (FMx-IRS), where the elements of the surface continuously change the phases of the incident signals. In this way, the FMx-IRS acts as a frequency mixer and makes the propagation environment nonlinear, thereby introducing new frequencies.
We study the basic features of the proposed concept and demonstrate its advantages that stem from the novel type of control over the wireless propagation.
The channel decoupling feature and the correlation between reflected channels are elaborated for the architecture, and are validated by the simulations.

\end{abstract}


\newtheorem{Definition}{Definition}
\newtheorem{Lemma}{Lemma}
\newtheorem{Theorem}{Theorem}
\newtheorem{Corollary}{Corollary}
\newtheorem{Proposition}{Proposition}
\newtheorem{Remark}{Remark}
\newtheorem{Property}{Property}
\newtheorem{Assumption}{Assumption}
\newcommand{\rl}[1]{\color{red}#1}

\newcommand{\edc}[1]{{\color{blue}{#1}}}
\newcommand{\rjw}[1]{{\color{Sepia}{\textit{\scriptsize #1}}}}
\newcommand{\e}[1]{\mathrm{e}^{#1}}

%
\IEEEpeerreviewmaketitle

\begin{spacing}{0.95}
\section{Introduction}

An intelligent reflecting surface (IRS)
consists of a large number of reflecting elements with tunable properties, which potentially can be used to unleash new operation modes and applications in wireless communications~\cite{Renzo2020b,8981888}.
In most of the existing literature, the reflecting elements are assumed to be phase shifters.
In a typical setting, the IRS can affect some of the dominant propagation paths between a user and a base station (BS). The objective is that the phases of the reflecting elements are chosen to create constructive interference at a desired location and thereby maximize the signal-to-noise ratio (SNR) of the received signal.
In order to achieve this type of beamforming at the IRS, it is necessary to obtain channel state information (CSI) regarding each reflected path. This is challenging when the IRS has a massive number of elements due to the huge pilot overhead \cite{Bjornson2020a}.

In order to overcome the problem, we propose to exploit an IRS in such a way that it acts as a \emph{frequency mixer} with a given frequency.
{{We are inspired by the time-domain digital-coding metasurface implemented in \cite{nsr/nwy135}, in which the surface element can generate different reflection amplitudes by a digital controller, which ranges from $-1$ to $1$ to represent the reflectivity changes between total inverse reflection and total reflection.
In our architecture, we {adjust the reflection amplitude according to a sinusoidal shape with a certain frequency uniquely associated with the surface in question.}
When a signal impinges on an element of the surface, it is manipulated by this time-variant reflector, such that it acts as a frequency mixer for the reflected signal.}}
This architecture, which we term \emph{Frequency-Mixing IRS (FMx-IRS)} makes the wireless propagation environment nonlinear and introduces new frequencies, not present in the signal sent from the transmitter.
With FMx-IRS, each FM operation at an IRS results in two new frequencies. For example, if a receiver obtains a signal through a direct path and through an FMx-IRS that pulsates with a single frequency, then the received signal has three frequencies: the original one through the direct path and two resulting from the FM operation at the IRS {that are symmetric around the original carrier frequency.}

This way of introducing non-linearity in the propagation environment can, potentially, open a plethora of new design opportunities. For example, when arriving at the receiver, the signals coming from the IRS can be uniquely identified by their carrier-frequency shifts.
This feature simplifies the channel estimation (CE) in IRS-aided systems,
which is challenging when the number of elements is very large. 
With FMx-IRS, the channels from each IRS can be estimated in parallel, potentially leading to a large decrease in the estimation overhead.

In this paper, we lay the foundation for FMx-IRS communications.
While at this point this is a rather speculative architecture, there are works~\cite{8901456} that have investigated the basis for physical implementation of this type of schemes. Our objective is to investigate it from a communication-theoretic viewpoint. Conceptually, FMx-IRS uses frequency mixers that are placed in the environment, rather than at the transmitter.
We start with the two-path channel model to introduce the basic principles behind FMx-IRS and describe the details of the operation.
Then we consider the infinite-path channel model to develop guidelines of choosing operating frequency at the surface.
We illustrate that channels from different frequency bands can be estimated in parallel and derive the upper bound of the achievable rate that can perfectly predict the system performance.
The numerical results indicate that by choosing proper operating frequencies, the reflected channels from the FMx-IRS are nearly independent and identically distributed.

\section{Basic Principles of FMx-IRS Operation}






In this section, we present the basic principles behind FMx-IRS in a simple example where a single-antenna user communicates with a single-antenna BS.
We adopt a continuous-time model throughout the paper where $t$ is the time variable. Furthermore, we consider continuous-time convolution and Fourier transforms.
We consider a single IRS that is equipped with a single frequency reconfigurable module (FRM). The phase manipulation at the FRM is time-variant with a certain frequency $f_r$.
The model can be generalized to multiple IRS, each with multiple FRMs.

\subsection{FRM-Based Operation}\label{FRMoperation}

To describe the operation of an FRM, let us assume that the user transmits a narrowband signal.
We assume the FRM in the IRS is able to change the amplitude and phase of the incoming narrowband signal according to a time-varying cosine function $\phi(t)$ with frequency $f_r$:
\begin{equation}
\phi \left( t \right) = \cos \left( {2\pi {f_r}t} \right).
\end{equation}
By letting $s(t)$ denote the incoming time-domain signal, the signal reflected by the FRM is
\begin{equation}\label{modular_r}
r(t)=s(t) \phi \left( t \right) = s(t) \cos \left( {2\pi {f_r}t} \right).
\end{equation}
Next, we express $s(t)$ as a function of the transmitted signal.
With a baseband signal ${x} = \left| {{x}} \right|{e^{j\theta }}$ where $\theta$ represents the phase, the radio frequency signal transmitted over the air becomes 
\begin{align}\label{r_rf}
{x_{{\text{R}}}}\left( t \right) = \Re \left\{ {x{e^{j2\pi {f_c}t}}} \right\} = \left|{{x}}\right|\cos \left( {2\pi {f_c}t}+\theta\right),
\end{align}
where $f_c$ is the carrier frequency. 
The channel from the user to the surface is a modeled as a single-path frequency-flat channel with impulse response $h(t) = \left| {{h}} \right| \delta \left( {t - {\tau _{{h}}}} \right)$ ($\delta(\cdot) $ is Dirac delta function), which remains constant during a symbol duration. The pathloss is $\left| {{h}} \right|$ and the time delay is ${\tau _{{h}}}$.
The signal $s(t)$ is the convolution of the input signal with the channel impulse response, 
so that the physical signal received at the IRS is
\begin{align}\label{r_re_IRS}
s\left( t \right) &= {\textstyle\int}_{-\infty}^{\infty}  {x_{{\text{R}}}}\left( {t - \tau } \right){h}\left( \tau  \right)d\tau    \notag \\
&= \left| {{h}} \right|\left| {{x}} \right|\cos \left( {2\pi {f_c}\left( {t - {\tau _{{h}}}} \right) + \theta } \right).
\end{align}
Using basic trigonometry, $r(t)$ in \eqref{modular_r} can be rewritten as
\begin{align}\label{r_refl_IRS}
r\left( t \right)& = {r_ + }\left( t \right) + {r_ - }\left( t \right) \notag \\
&= \tfrac{1}{2}\left| {{h}} \right|\left| {{x}} \right|\cos \left( {2\pi \left( {{f_c} + {f_r}} \right)t - 2\pi {f_c}{\tau _{{h}}} + \theta } \right) \notag \\
& + \tfrac{1}{2}\left| {{h}} \right|\left| {{x}} \right|\cos \left( {2\pi \left( {{f_c} - {f_r}} \right)t - 2\pi {f_c}{\tau _{{h}}} + \theta } \right).
\end{align}
Note that the signal reflected from the IRS appears now at two different carrier frequencies: $f_c + f_r$ and $f_c - f_r$.



We now study the link from the FMx-IRS to the BS, which we also assume is a single-path frequency-flat channel. Hence, the impulse response can be expressed as
$g\left( t \right)
= |g| \delta \left( {t \!-\! {\tau_{g}}} \right)$, where ${g}$ and $\tau_{g}$ represent the pathloss and the time delay.
After convolution with the channel $g(t)$, we obtain 
\begin{align}\label{s_refl_IRS}
w\left( t \right) = {\left| {{g}} \right|{r_ + }\left( {t - {\tau _{g}}} \right)}  + {\left| {{g}} \right|{r_ - }\left( {t - {\tau _{g}}} \right)}.
\end{align}
We will now take the direct path between the user and BS into account, which has impulse response $h_{\rm d}(t) = |h_{\rm d}|\delta(t - \tau_{\rm d})$ with $|h_{\rm d}|$ being the pathloss and $\tau_{\rm d}$ the time delay. The received signal at the BS is then given as
\begin{align}\label{Y_pm}
&y\left( t \right) \!= \!\left| {{h_{\text{d}}}} \right|\!\left| {{x}} \right|\cos\! \left( {2\pi {f_c}t\! - \!2\pi {f_c}{\tau _{\text{d}}}\! +\! \theta } \right) \!+\! w\left(t\right).
\end{align}

At the BS, the signal is demodulated to baseband and low-pass filtered. With the in-phase and quadrature demodulation at $f_c$, the signal has the following form:
\begin{align}\label{Y_de}
{y_{{\text{de}}}}\left( t \right) = y\left( t \right)\cos \left( {2\pi {f_c}t} \right) + jy\left( t \right)\sin \left( {2\pi {f_c}t} \right).
\end{align} \rjw{}
After a low-pass filter, the output of (\ref{Y_de}) is
\begin{align}\label{Y_delp}
y_{{\text{de}}}^{{\text{lp}}}\left( t \right) &= \left| {{h_{\text{d}}}} \right|\left| {{x}} \right|{e^{j\left( { - 2\pi {f_c}{\tau _{\text{d}}} + \theta } \right)}}\notag \\
&+ \tfrac{1}{2}\left| h \right|\left| {{x}} \right|{\left| {{g}} \right|{e^{j\left( {2\pi {f_r}t - 2\pi \left( {{f_c} + {f_r}} \right){\tau _{g}} - 2\pi {f_c}{\tau _h} + \theta } \right)}}}  \notag \\
&+ \tfrac{1}{2}\left| h \right|\left| {{x}} \right|{\left| {{g}} \right|\!{e^{j\left( { - 2\pi {f_r}t - 2\pi \left( {{f_c} \!- \!{f_r}} \right){\tau _{g}} \!-\! 2\pi {f_c}{\tau _h} + \theta } \right)}}}.
\end{align}
It is represented in the frequency domain by
\begin{align}\label{Y_delp_fre}
Y(f) &= {h_{\text{d}}}{x}\delta \left( f \right)\notag\\
&+ \tfrac{1}{2}h{x}{{g_{+ }}\delta \left( {f - {f_r}} \right)}
+ \tfrac{1}{2}h{x} {{g_{- }}\delta \left( {f + {f_r}} \right)} ,
\end{align}
where ${h_{\rm{d}}} = |h_{\rm d}|e^{-j2\pi f_c \tau_{\rm d}}$, ${h} = |h|e^{-j2\pi f_c \tau_{h}}$, and ${g_ {\pm}} = {\left| {{g}} \right|{e^{ - j2\pi \left( {{f_c} \pm {f_r}} \right){\tau _{g}}}}} $.
Note that although the time delay is the same for the channels at $f_r$ and $-f_r$, the phase shifts are different due to the modulation by different carrier frequencies.
Since the three components in (\ref{Y_delp_fre}) are separable in the frequency domain, we can  extract the signals from the direct path and the reflected paths from the IRS:
\begin{align}\label{Y_de_freq}
\left\{ {\begin{array}{*{20}{l}}
  {{y_{{\text{d}}}} = h_{\rm d}x \delta(f)},\\
  {{y_{+}} = z_+x \delta(f-f_r)}, \\
  {{y_{-}} = z_-x \delta(f+f_r)},
\end{array}} \right.
\end{align}
where $z_{\pm} = \tfrac{1}{2}hg_{\pm}$. We {denote this as a \emph{decoupling feature}.}

This feature of an FMx-IRS system is not only appearing in the two-path scenario, but also exist in scattering environments since the frequency of every path reflected by the FRM is shifted by $\pm f_r$.
Therefore, we can always observe two components at new frequencies.
However, in a rich scattering environment, reflected paths from FMx-IRS may have some new features which is analyzed in Section \ref{Correlation_Analysis}.

\begin{Remark}
Different from the conventional two-path interference model,
the pathloss of the resulting channel is  based on the phase difference of two interfering rays.
With FMx-IRS, the direct and reflected signals from the surface are separated at different frequencies and the pathloss does not fluctuate according to the phase difference of the two signals. {Nevertheless, the received signal has a larger bandwidth compared to the original one.}
\end{Remark}

\subsection{Correlation Analysis}\label{Correlation_Analysis}

We now consider a more general propagation environment: a rich scattering environment where the channel of each link consists of a large number of paths.
We first elaborate that all channels asymptotically follow a complex Gaussian distribution  and then illustrate the correlation between the two reflected channels at the new frequencies.

Assume that the normalized impulse responses of the channel from the user to BS $h(t)$, the user to the surface $h_{\rm{d}}(t)$ and the surface to the BS $g(t)$ can be decomposed as the sum of $I$, $N$, and $L$ propagation paths, respectively:
\begin{align}
\label{h}&h\left( t \right)
= \sum\nolimits_{i=1}^{I}
|h_i| \delta \left( {t - {\tau_{h,i}}} \right),\\
\label{h_d}&h_{\rm{d}}\left( t \right)
= \sum\nolimits_{n=1}^{N}
|h_{{\rm{d}},n}| \delta \left( {t - {\tau_{h_{\rm{d}},n}}} \right),\\
\label{g}&g\left( t \right)
= \sum\nolimits_{l=1}^{L}|g_l| \delta \left( {t - {\tau_{g,l}}} \right),
\end{align}
where ${h_i}$, $h_{{\rm{d}},n}$, and ${g_l}$ represent the amplitude, and ${\tau_{h,i}}$, ${\tau_{h_{\rm{d}},n}}$ and $\tau_{g,l}$ represent the corresponding delays.
Since the transmission of a narrowband signal is considered in the paper, the symbol interval is assumed to be much larger than the maximum delay of all paths and channels are assumed to be constant during a symbol interval.

Applying the same derivations in Section \ref{FRMoperation}, the baseband equivalent channels in (\ref{Y_delp_fre}) can then be expressed as $h = \sum\nolimits_{i = 1}^I {{h_i}}$, ${h_{\text{d}}} = \sum\nolimits_{n = 1}^N {{h_{{\text{d}},n}}}$, and
\begin{align}
\label{g_multi}
{g_ \pm } = \sum\nolimits_{l = 1}^L {{g_{ \pm ,l}}}
\end{align}
in the frequency domain, where ${h_i} = \left| {{h_i}} \right|{e^{ - j2\pi {f_c}{\tau _{h,i}}}}$,
${h_{{\text{d}},n}} = \left| {{h_{{\text{d}},n}}} \right|{e^{ - j2\pi {f_c}{\tau _{{h_{\text{d}}},n}}}}$, and ${g_{ \pm ,l}} = \left| {{g_l}} \right|{e^{ - j2\pi \left( {{f_c} \pm {f_r}} \right){\tau _{g,l}}}}$.


\begin{Lemma}\label{E_cossin}
Suppose $\tau$ follows a uniform distribution with probability density function (PDF) $f_{\tau}(\tau) = \tfrac{1}{D}$ ranging from $0$ to $D$. For any scalar $a$, we have following expectations:
\begin{align}
\label{E_cos}&{\mathsf{E}}_\tau\left\{ {\cos \left( {2\pi a\tau } \right)} \right\} = \tfrac{{\sin \left( {2\pi aD} \right)}}{{2\pi aD}}, \\
\label{E_sin}&{\mathsf{E}}_\tau\left\{ {\sin \left( {2\pi a\tau } \right)} \right\} = \tfrac{{1 - \cos \left( {2\pi aD} \right)}}{{2\pi aD}}, \\
\label{E_cos2}&{\mathsf{E}}_\tau\big\{ {\cos {{\left( {2\pi a\tau } \right)}^2}} \big\} = \tfrac{1}{2} + \tfrac{{\sin \left( {4\pi aD} \right)}}{{8\pi aD}},\\
\label{E_sin2}&{\mathsf{E}}_\tau\big\{ {\sin {{\left( {2\pi a\tau } \right)}^2}} \big\} = \tfrac{1}{2} - \tfrac{{\sin \left( {4\pi aD} \right)}}{{8\pi aD}}.
\end{align}
\end{Lemma}

From (\ref{g_multi}), we can rewrite $g_{\pm}$ as
\begin{align}\label{g_cossin}
{g_{ \pm}} &= \sum\nolimits_{l=1}^{L}\left| {{g_l}} \right|\cos \left( {2\pi \left( {{f_c}\! \pm \!{f_r}} \right){\tau _{g,l}}} \right)\! \notag\\
&- j\sum\nolimits_{l=1}^{L}\left| {{g_l}} \right|\sin \left( {2\pi \left( {{f_c} \!\pm \!{f_r}} \right){\tau _{g,l}}} \right).
\end{align}
Let $\tau_{\max}$ be the maximum delay among all paths. 
It is reasonable to assume that the delays of $L$ paths are uniformly distributed in the range $\left[0,\tau_{\max}\right]$.
Therefore, according to Lemma \ref{E_cossin}, since ${{f_c} \pm {f_r}} \gg (2\pi\tau_{\max})^{-1}$, the means of the real part and the imaginary part of ${g_{ \pm ,l}}$ are approximately equal to $0$, and their variance approximately equal to $1/2$.
Along with the fact that $\mathsf{E}\{|g_{l}|^2\} = 1/L$, and according to the {Central Limit Theorem}, when $L$ goes to infinity, the distribution of ${g_ \pm }$
tends toward a complex normal distribution, i.e., ${g_ \pm } \sim \mathcal{CN}\left( {0,1} \right)$.
Similarly, we can obtain that $h\sim \mathcal{CN}\left( {0,1} \right)$ and $h_{\rm d}\sim \mathcal{CN}\left( {0,1} \right)$ using the same method.

The channel $h$ and $h_{\rm{d}}$ are mutually independent, and are independent with $g_{\pm}$ by nature due to the different propagation environments.
However, since ${g_+ }$ and ${g_-}$ share the same amplitude and the delay, it is necessary to evaluate the correlation between the two reflected channels.

\begin{Proposition}
The correlation $\rho({f_r})=\left| {{\mathsf{E}_{{g_l},{\tau _{g,l}}}}\left\{ {{g_ + }g_ - ^H} \right\}} \right|$ with respect to the frequency shift $f_r$ is given as
\begin{align}
\rho({f_r}) =\tfrac{{\sqrt {2 - 2\cos \left( {4\pi {f_r}{\tau _{\max }}} \right)} }}{{4\pi {f_r}{\tau _{\max }}}}.
\end{align}
\end{Proposition}
\begin{IEEEproof}
Noting that the $L$ paths in $g_{\pm}$  are mutual independent as they involve independent random variables,
the correlation $\rho({f_r})$ is
\begin{align}\label{correlation_E}
\rho({f_r}) &=\Big|\mathsf{E}_{{g_l},{\tau _{g,l}}}\Big\{ \Big( \sum\nolimits_{l = 1}^L
{ {g_{ + ,l}}  }\Big) \Big(\sum\nolimits_{l' = 1}^L {g_{ - ,l'}^H}  \Big)  \Big\} \Big| \notag \\
&= \Big| {{\mathsf{E}_{{g_l},{\tau _{g,l}}}}\Big\{ {\sum\nolimits_{l = 1}^L {{g_{ + ,l}}g_{ - ,l}^H}   } \Big\}}
\Big|.
\end{align}
Hence, by observing that $g_l$ and $\tau_{g,l}$ are mutually independent with $\mathsf{E}\{|g_{l}|^2\} = 1/L$, we can compute (\ref{correlation_E}) as
\begin{align}\label{correlation_int}
\rho({f_r})& =L{\mathsf{E}}\big\{ {{{\left| {{g_l}} \right|}^2}} \big\}\left| {{\mathsf{E}}_{\tau_{g,l}}\left\{ {{e^{ - j4\pi {f_r}\tau_{g,l} }}} \right\}} \right|  \notag \\
&\mathop  = \limits^{\left( a \right)}  \sqrt {{{\left( {\tfrac{{\sin \left( {4\pi {f_r}{\tau _{\max }}} \right)}}{{4\pi {f_r}{\tau _{\max }}}}} \right)}^2} + {{\left( {\tfrac{{1 - \cos \left( {4\pi {f_r}{\tau _{\max }}} \right)}}{{4\pi {f_r}{\tau _{\max }}}}} \right)}^2}}  \notag\\
& =\tfrac{{\sqrt {2 - 2\cos \left( {4\pi {f_r}{\tau _{\max }}} \right)} }}{{4\pi {f_r}{\tau _{\max }}}},
\end{align}
where $\left( a \right)$ is obtained using (\ref{E_cos}) and (\ref{E_sin}).
\end{IEEEproof}

Proposition 1 provides a guideline for the selection of $f_r$. Suppose we select $f_r = i\Delta f$ where $\Delta f = 1/(2\tau_{\max})$ and $i$ is a positive integer. We then obtain two uncorrelated channels $g_{+}$ and $g_{-}$ at new frequencies.

\subsection{Decoupling Feature}

\begin{figure}[t!]
  \centering
  \includegraphics[width=8.5cm]{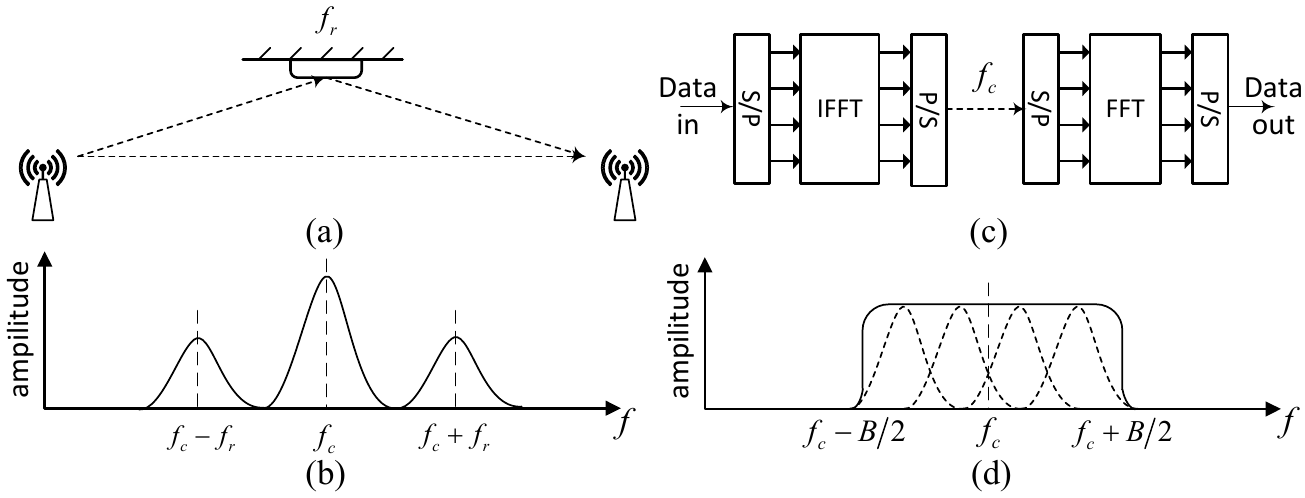}
  \caption{A simple example of an FMx-IRS-aided system is shown in (a), where a single-antenna transmitter communicates with a single-antenna receiver aided by an FRM operating on frequency $f_r$; (b) shows the corresponding frequency response of the received signal;
  (c) give an example of a wideband system with carrier at $f_c$ and bandwidth $B$, and (d) shows the corresponding frequency response.}
  \label{Frequency_domain}
\end{figure}

As illustrated in Fig.~{\ref{Frequency_domain}}(b), the frequency response of the received signal consists of three components with two of them around $f_c \pm f_{r}$.
Since the channels are decoupled in the frequency domain, one pilot can estimate all three channels, which we call the channel decoupling feature.
This is significantly different from a conventional IRS system, where one element generates one channel, and the reflective channel is coupled with the channel from the direct path at the receiver. To separate the channel of each link, it is required to send two pilot symbols \cite{9053695}. 
The decoupling feature is particularly important since, in conventional IRS systems, the pilot overhead required for estimating each link grows linearly with the number of the elements on the surface, which results in a huge degradation of achievable rate under user mobility.
However, in an FMx-IRS aided system, each of the FRMs can be associated to an unique frequency. As long as the frequency spacing is larger than the signal bandwidth, all the channels reflected by the surface are separable in the frequency domain, thereby {sparing the use of pilot sequences.}

Moreover, the FMx-IRS-aided system shows an essential different feature compared with a conventional wideband system shown in Fig.~{\ref{Frequency_domain}}(c).
In a conventional wideband system, the power of pilot and data are allocated across the bandwidth $B$, whereas in an an FMx-IRS-aided system, the transmission power is concentrated in a narrow band, thereby observing a much higher SNR at the receiver side.

\section{General System Model}






We will now consider a general model in uplink direction where one BS with $M$ antennas serves one single-antenna user aided by one FMx-IRS.
We assume that user is randomly located in area while the BS and the surface have fixed positions.
The surface is divided into an arrany of $V \times S$ FRMs, and the $(v,s)$th FRM carries out a frequency modulation operation at frequency $f_{v,s}$.
We assume the frequency spacing between two adjacent frequency shifts is $f_n$, and $f_{v,s} = ((v-1)S +s) f_n$.
Therefore, a single narrowband signal from the transmitter will result in a set of $2SV+1$ received narrowband signals at the receiver, and from those signals we can uniquely estimate the channel response from each FRM.

In this section, all the quantities are in the frequency domain, so that we omit the frequency variable $f$ for simplicity.
We denote by ${\mathbf{h}}_{{\rm{d}}}\in\mathbb{C}^{M\times 1}$ and ${{h}}_{v,s}\in\mathbb{C}$ the direct channel from user to the BS and the channel from user to the $(v,s)$th FRM of the surface, respectively.
Regarding the paths from the surface to the BS, each FMx-IRS creates two channels at two different frequency bands.
We denote by $\mathbf{g}_{+,v,s}\in\mathbb{C}^{M\times 1}$ and $\mathbf{g}_{-,v,s}\in\mathbb{C}^{M\times 1}$ the channels from the $(v,s)$th FRM to the BS at frequency $f_c \pm f_{v,s}$.



We assume that $|x|=1$ is a complex signal variable that can represent either a pilot or data symbol.
The frequency-domain received signal ${\mathbf{y}}\in\mathbb{C}^{(2VS+1)M \times 1}$ is then written as 
\begin{align}\label{r_l_simp}
{\mathbf{y}} = \sqrt {p}  {{\mathbf{h}}_{{\rm{all}}}}{x} + {\mathbf{n}},
\end{align}
where $p$ is the power of each symbol,
\begin{align}
\label{y_mtrix}
&{\mathbf{y}} = {\left[ {{\mathbf{y}}_{{\text{d}}}^T,{\mathbf{y}}_{ + ,1,1}^T, {\mathbf{y}}_{ - ,1,1}^T, \cdots ,{\mathbf{y}}_{ + ,V,S}^T,{\mathbf{y}}_{ - ,V,S}^T} \right]^T},\\
\label{H_mtrix}
&{{\mathbf{h}}_{{\text{all}}}} = {\left[ {{\mathbf{h}}_{{\text{d}}}^T,{\mathbf{z}}_{ + ,1,1}^T, {\mathbf{z}}_{ - ,1,1}^T, \cdots ,{\mathbf{z}}_{ + ,V,S}^T ,{\mathbf{z}}_{ - ,V,S}^T} \right]^T}
\end{align}
with ${{\mathbf{z}}_{ \pm ,v,s}} = \tfrac{1}{2}{{{h}}_{v,s}}{{\mathbf{g}}_{ \pm ,v,s}}$ representing the cascaded channel from the user to the $(v,s)$th FRM at surface to the BS.
$\mathbf{n}\in\mathbb{C}^{(2VS+1)M\times 1}$ is AWGN with i.i.d. $\mathcal{CN}(0, 1)$ elements.

\section{Channel Estimation and Downlink Precoder Design In FMx-IRS-Aided Systems}\label{Channel Estimation}


In the section, we elaborate the CE procedure in the different propagation environments, and design the precoder for infinite-path channel model to improve the achievable SNR.

\subsection{CE in Two-Path Model}



For this model, the pathloss and the delay depend on the distance of the propagation path.
We assume that the 3D coordinate of the user, the first BS antenna, and the $(1,1)$th FRM at the surface are $\mathbf{o}_{\rm{u}}=(b_{\rm u},e_{\rm u},u_{\rm u})$, $\mathbf{o}_{\rm{b},1}=(b_{\rm b},e_{\rm b},u_{\rm b})$ and $\mathbf{o}_{\rm{s},1,1}=(b_{\rm s},e_{\rm s},u_{\rm s})$, respectively. Therefore, \rjw{the coordinates of} the $m$th BS antenna and the $(v,s)$th FRM at the surface are given by $\mathbf{o}_{{\rm{b}},m} =(b_{\rm b}+(m-1)d_{\rm b},e_{\rm b},u_{\rm b})$ and $\mathbf{o}_{{\rm{s}},v,s} =(b_{\rm s},e_{\rm s}+(v-1)d_{\rm s},u_{\rm s}+(s-1)d_{\rm s})$ with $d_{\rm b}$ and $d_{\rm s}$ being the antenna spacing at the BS and the FRM spacing at the surface.

Using these coordinates, the channel components of the user-surface link, user-BS link and surface-BS link can be modelled as
\begin{align}
\label{h_vs}{h_{v,s}} &=\left| {{h_{v,s}}} \right| {e^{ - j{\kappa}\left\| {{{\mathbf{o}}_{\text{u}}} - {{\mathbf{o}}_{{\text{s}},v,s}}} \right\|}}, \\
\label{h_dm}{h_{\text{d},m}} &= \left| {{h_{{\text{d,}}m}}} \right|{e^{ - j{\kappa}\left\| {{{\mathbf{o}}_{\text{u}}} - {{\mathbf{o}}_{{\text{b}},m}}} \right\|}}, \\
\label{g_vsm}{g_{ \pm ,v,s,m}} &=\left| {{g_{v,s,m}}} \right|{e^{ - j{\kappa _{\pm,v,s}}\left\| {{{\mathbf{o}}_{{\text{b}},m}} - {{\mathbf{o}}_{{\text{s}},v,s}}} \right\|}},
\end{align}
where $\left| {{h_{v,s}}} \right| = {({\left\| {{{\mathbf{o}}_{\text{u}}} - {{\mathbf{o}}_{{\text{s}},v,s}}} \right\|}/{d_0})^{ - \alpha }}$,
$\left| {{h_{{\text{d,}}m}}} \right|$ $={(\left\| {{{\mathbf{o}}_{\text{u}}} - {{\mathbf{o}}_{{\text{b}},m}}} \right\|/{d_0})^{ - \alpha }}$,
and $\left| {{g_{v,s,m}}} \right|= {(\left\| {{{\mathbf{o}}_{{\text{b}},m}} - {{\mathbf{o}}_{{\text{s}},v,s}}} \right\|/{d_0})^{ - \alpha }}$ with $d_0$ representing the reference distance according Friis transmission formula;
$\kappa = \tfrac{2\pi f_c}{c}$ and $\kappa_{\pm,v,s} = \tfrac{2\pi(f_c \pm f_{v,s})}{c}$ with $c$ being the light speed; $\alpha$ represents the pathloss exponent with typical value $\alpha=2$ representing the LoS scenario. Therefore, the cascaded channel is given as
\begin{align}
\label{z_vsm}{z_{ \pm ,v,s,m}} = |{z_{v,s,m}}|{e^{ - j({\kappa}\left\| {{{\mathbf{o}}_{\text{u}}} - {{\mathbf{o}}_{{\text{s}},v,s}}} \right\| + {\kappa _{\pm,v,s}}\left\| {{{\mathbf{o}}_{{\text{b}},m}} - {{\mathbf{o}}_{{\text{s}},v,s}}} \right\|)}},
\end{align}
where $|{z_{v,s,m}}| = |\tfrac{1}{2}h_{{v,s}}{g_{ v,s,m}}|$.
Note that the channel of the surface-BS link is deterministic due to the fixed position of the surface and the BS.

Since $\mathbf{h}_{\rm d}$, $\mathbf{z}_{\pm,s,v}$ are decoupled in the frequency domain, their estimation can be performed in parallel.
Adopting the conventional least squares estimator, we obtain the channel estimates as
\begin{align}\label{single_path_h_all_est}
\hat{\mathbf{h}}_{\text{all}} = \frac{1}{\sqrt{p}}x^*{\mathbf{y}},
\end{align}
where $x^*$ represents the conjugate transpose of $x$.
The estimation error $\tilde{\mathbf{h}}_{\text{all}}=\hat{\mathbf{h}}_{\text{all}} -{\mathbf{h}}_{\text{all}}$ has i.i.d.~$\mathcal{CN}(0, 1/p)$ elements.


\subsection{CE in Infinite-Path Model}\label{CE_inf}

As illustrated in Section~\ref{Correlation_Analysis}, we obtain ${{h}}_{{\rm{d}},m}\sim \mathcal{CN}({0},1)$ and ${{h}}_{v,s}\sim \mathcal{CN}({0},1)$ when the number of paths grow large.
Regarding the channels from the surface to the BS, instead of simply assuming they are complex Gaussian value, it is required to model each of them as the sum of $L$ ($L\to \infty$) paths:
\begin{align}
\label{g_vsm_multi}{g_{ \pm ,v,s,m}} = \sum\limits_{l = 1}^L {{g_{ \pm ,v,s,m,l}}} ,
\end{align}
where ${g_{ \pm ,v,s,m,l}} = \left| {{g_{v,s,m,l}}} \right|{e^{ - j2\pi \left( {{f_c} \pm {f_{v,s}}} \right){\tau _{v,s,m,l}}}}$, since ${g_{ + ,v,s,m}} $ and ${g_{ - ,v,s,m}} $ are not mutually independent. Moreover, we model the  amplitude and the delay of the $l$th path randomly as $\mathsf{E}\{\left| {{g_{v,s,m,l}}} \right|^2\} = 1/L$ and $\tau _{v,s,m,l}$ uniformly distributed in $[0,\tau_{\max}]$. 
In this manner, ${g_{ + ,v,s,m}}$ and ${g_{ - ,v,s,m}} $ follow a complex Gaussian distribution with the correlation being $\rho({f_{v,s}})$.
Therefore, the cascaded channel $z_{\pm, v,s,m} =\tfrac{1}{2} h_{{v,s}}{g_{ \pm ,v,s,m}}$ is then the product of two complex Gaussian variables, and follows a complex product-normal distribution \cite{NADARAJAH2016201}, which is symmetric around the origin with the variance being $1/4$.


We apply a minimum mean square error (MMSE) estimation approach to estimate $\mathbf{h}_{\rm{all}}$.
Noting that the estimation of $\mathbf{h}_{\rm d}$, $\mathbf{z}_{\pm,v,s}$ are decoupled. We obtain the MMSE estimator as  \cite{Kay1993}:
\begin{align}\label{H_d_estimate_exp}
{{{\mathbf{\hat h}}}_{\text{d}}} &= \tfrac{1}{{\sqrt p }}{{\mathbf{y}}_{\text{d}}}x^*{\left( {1 + 1/p} \right)^{ - 1}},\\
\label{Z_pm_estimate}
{{{\mathbf{\hat z}}}_{ \pm ,v,s}} &= \tfrac{1}{{\sqrt p }}{{\mathbf{y}}_{ \pm ,v,s}}x^*{\left( {4 + {1}/{p}} \right)^{ - 1}}
\end{align}
by realizing that the covariance matrix of LoS channel and the cascaded channel are $\mathbf{I}_M$ and $\tfrac{1}{4}\mathbf{I}_M$, respectively.
Let ${{{\mathbf{\tilde h}}}_{\text{d}}} = {{{\mathbf{\hat h}}}_{\text{d}}} - {{\mathbf{h}}_{\text{d}}} $ and ${{{\mathbf{\tilde z}}}_ {\pm,v,s}  }={{{\mathbf{\hat z}}}_ {\pm,v,s} } - {{\mathbf{z}}_ {\pm,v,s}} $ being the channel estimation error vector for the direct path and reflected paths at $f_c \pm f_s$, respectively.
The variance of elements of estimation error for direct path and reflect path are equal to
$\mathsf{var}\{ { {{\tilde h}}}_{\text{d},m} \} = \tfrac{1}{{1 + {p}}}$, and $\mathsf{var}\{  {{{\tilde z}}_{\pm,v,s,m}} \} = \tfrac{1}{{1 + (1/4){p}}}$, respectively \cite{Kay1993}.
Equations (\ref{single_path_h_all_est}), (\ref{H_d_estimate_exp}) and (\ref{Z_pm_estimate}) indicate that the pilot-based channel estimation scheme is feasible for estimating the cascaded channel for both two-path model and multipath model without requiring extra pilot overhead.

\subsection{Rate Analysis for Infinite-path Model}

Assuming that perfect CSI is available at the BS, the achievable rate can be written
in the following form:
\begin{align}\label{Erg_capacity}
R_{\rm p} = \mathsf{E}_{\mathbf{h}_{\rm{all}}} \left[\log_2(1 + p\mathbf{h}^H_{\rm{all}}\mathbf{h}_{\rm{all}})\right].
\end{align}
An analytical evaluation of (\ref{Erg_capacity}) is complicated due to the different distribution of the direct channel and cascaded channel.
However, using Jensen’s inequality, we can compute and upper bound on the achievable rate in closed form:
\begin{align}
{R_{\text{p}}} \leqslant {\log _2}\left( {1 + p{\mathsf{E}_{{{\mathbf{h}}_{{\text{all}}}}}}\left[ {{\mathbf{h}}_{{\text{all}}}^H{{\mathbf{h}}_{{\text{all}}}}} \right]} \right).
\end{align}
Based on the statistic properties given in Section\,\ref{CE_inf}, it follows that
\begin{align}
{\mathsf{E}_{{{\mathbf{h}}_{{\text{all}}}}}}\left[ {{\mathbf{h}}_{{\text{all}}}^H{{\mathbf{h}}_{{\text{all}}}}} \right] = M + \frac{{MSV}}{2}.
\end{align}
\begin{Proposition}
With perfect CSI, the {achievable rate} for the FMx-IRS-aided system is upper bounded by
\begin{align}
{R_{\text{p}}} \leqslant {\log _2}\left( {1 + pM\left( {1 + \frac{{SV}}{2}} \right)} \right).
\end{align}
\end{Proposition}
The result indicates that the capacity grows linearly with the number of FRMs on the surface.

\section{Numerical Results}




In the simulation, we set the coordinates of the first BS antenna and the $(1,1)$ FRM at the surface at $\mathbf{o}_{{\rm b},1}=(30, 30, 10)$\,m and $\mathbf{o}_{{\rm s},1,1}=(0, 0, 4)$\,m, respectively. The antenna spacing and FRM spacing is set as
$d_{\rm b}=0.1$\,m, $d_{\rm s}=0.1$\,m, while the reference distance is set as $d_0 = 50$\,m.

\begin{figure}[b!]
  \centering
  \includegraphics[width=8cm]{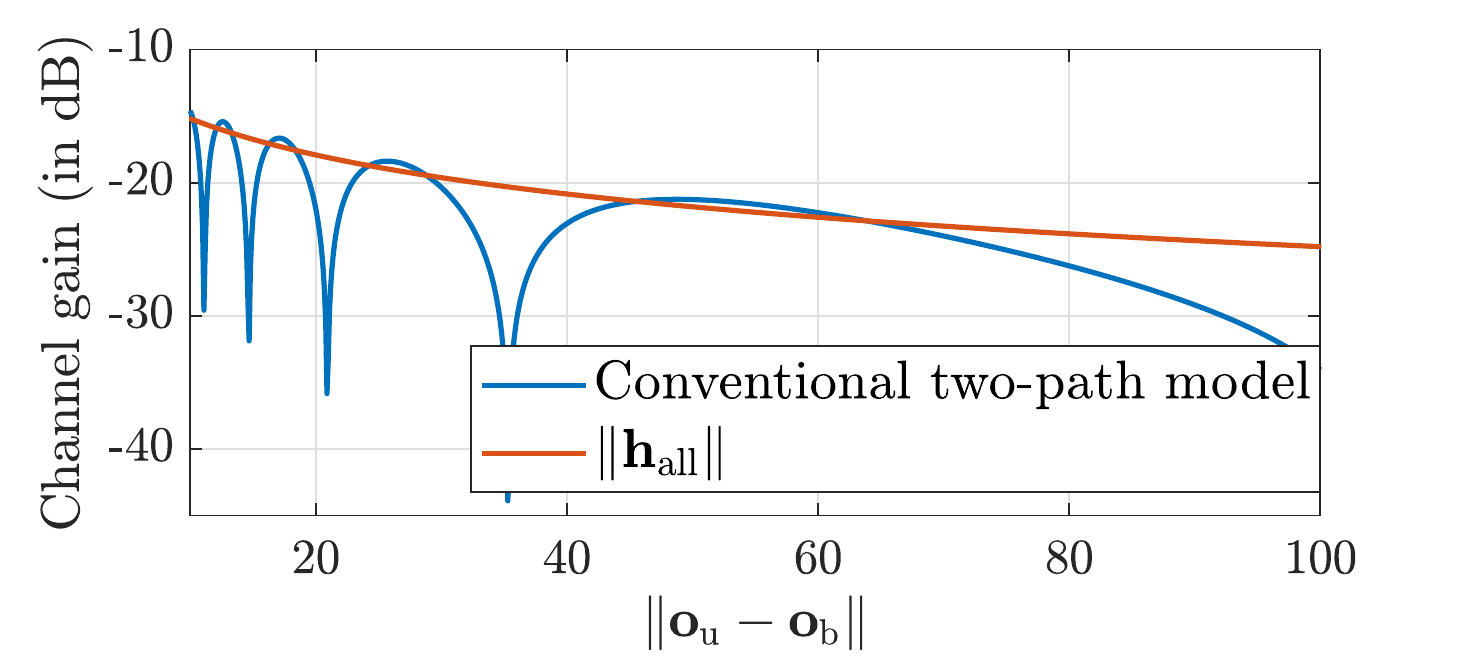}
  \caption{Pathloss v.s. distance between user and BS under two-path model. The result is shown for $V,S,M =1$ and $f_n = 0.1 \Delta f$.}
  \label{PL_distance}
\end{figure}

\begin{figure}[!t]
  \centering
  \includegraphics[width=8.8cm]{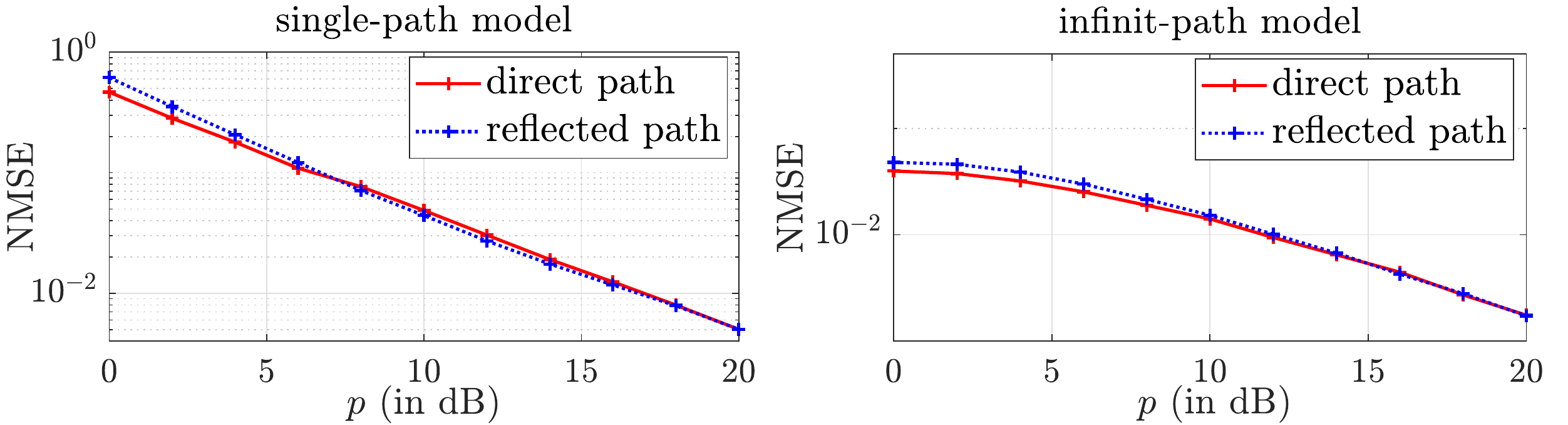}
  \caption{Channel NMSE v.s. the transmit power for different channel model. The results are shown for $V,S=2$, and $\mathbf{o}_{{\rm u}}=(-50, 30, 1)$\,m.}
  \label{Simulation_1}
\end{figure}

\begin{figure}[!t]
  \centering
  \includegraphics[width=8.8cm]{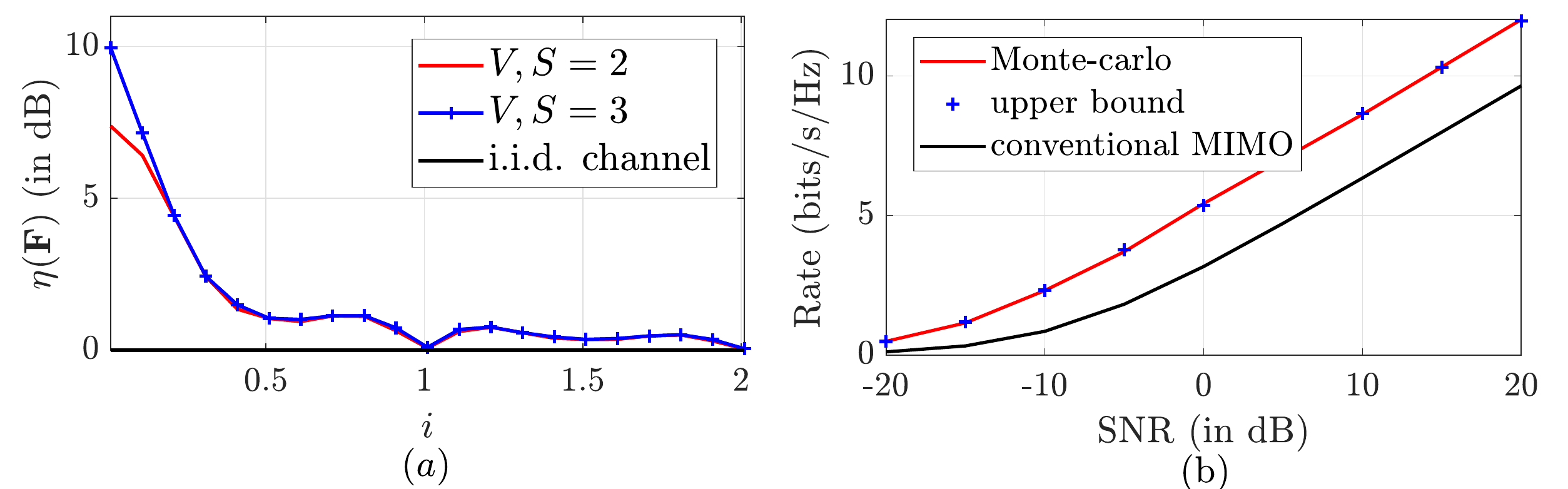}
  \caption{Impact of correlation among reflected paths under infinite-path model, in which (a) shows the conditioning number of $\mathbf{F}$ w.r.t $i = f_n/\Delta f$; (b) compares the capacity upper bounds with real capacity where $f_n = \Delta f$, $V= S =2$, and $M=8$.}
  \label{Simulation_2}
\end{figure}

{Fig.~\ref{PL_distance} compares the channel gain of the classical two-path model and the two-path model in which one of the paths goes through an FMx-RIS. The fluctuations in the classical two-path model are due to the superposition of the two rays with different phases. Following Remark 1, FMx-RIS decouples the two paths in frequency and avoids this superposition, which stabilizes the channel gain of the received signal as a function of distance, at the expense of a larger occupied bandwidth.}

Fig.~\ref{Simulation_1} exhibits the normalized MSE (NMSE) of channel estimation under  different propagation environment.
For the two-path model, the estimation of the direct path and reflected path gives the same performance, while under the infinite-path model, the estimation performance of the directed path outperforms that of the reflected path at low-SNR.

In Fig.~\ref{Simulation_2}(a), we illustrate the condition number $\eta(\mathbf{F})$ which equals to the ratio between the maximum singular value and the minimum singular value of $\mathbf{F}$.
We observe that when $i$ is an integer, the condition number is 0\,dB, indicating that $\mathbf{F}$ is an identity matrix as expected.
Moreover, when the frequency spacing is small, the more FRMs the surface contain, the closer $\mathbf{F}$ is to being singular.
Fig.~\ref{Simulation_2}(b) compares the tightness of the analytical rate upper bound compare with the real capacity.
We also include the performance of the conventional MIMO as the benchmark.
It can be observed that the upper bound converge to the real achievable rate really tight which indicates that analytical result can perfectly predict the rate performance. Moreover, we can obtain rate gain compared with conventional MIMO which is intuitive due to the extra contribution from the reflected paths.

\section{Conclusion}

In the paper, we proposed a novel FMx-IRS-aided system, where the IRS consists of several number modules that are capable of manipulating the frequencies of the incident signals.
We first demonstrated the basic principles of the FMx-IRS architecture and illustrated its advantages in terms of channel decoupling.
Then, we provided guidelines for choosing the operating frequencies for the infinite-path channel model.
The channel estimation is investigated for the different propagation environments, and the upper bound of the achievable rate is derived for the infinite-path channel model.
The numerical results indicate the tightness of rate bound, and with the appropriate operating frequencies, FMx-IRS generates a nearly i.i.d.~propagation environment.
{Clearly, the concept of FMx-IRS challenges the present notion of frequency occupancy determined by the transmitter and can have consequences for spectrum allocation and interference management, which is an interesting venue for future work.}

\end{spacing}

\ifCLASSOPTIONcaptionsoff
  \newpage
\fi



%
%
%

\footnotesize
\bibliographystyle{IEEEtran}

%

%
%
%




\end{document}